\documentclass[12pt]{article}

\usepackage{graphicx,amssymb,url,mathrsfs, amsmath}
\usepackage{eucal}
\usepackage{wrapfig}
\usepackage{boxedminipage}
\usepackage{setspace}
\usepackage{subfigure}

\def\a  {\alpha}                
       \def\d  {\delta}        
\def\e  {\epsilon}        \def\k  {\kappa}
\def\l  {\lambda}             \def\m  {\mu}
\def\n  {\nu}          \def\s  {\sigma}        
\def\t  {\tau}                 
   \def\w  {\omega}

\newcommand{\cala}{\mbox{${\cal A}$}} 
 
 \newcommand{\calf}{\mbox{${\cal F}$}}

\newcommand{\calq}{\mbox{${\cal Q}$}} 
 
 \newcommand{\calv}{\mbox{${\cal V}$}}

\newcommand{\scra}{\mbox{${\mathscr A}$}}

\def\IR{{\hbox{{\rm I}\kern-.2em\hbox{\rm R}}}}
\def\IB{{\hbox{{\rm I}\kern-.2em\hbox{\rm B}}}}
\def\IN{{\hbox{{\rm I}\kern-.2em\hbox{\rm N}}}}
\def\IC{\,\,{\hbox{{\rm I}\kern-.59em\hbox{\bf C}}}}
\def\IZ{{\hbox{{\rm Z}\kern-.4em\hbox{\rm Z}}}}
\def\IP{{\hbox{{\rm I}\kern-.2em\hbox{\rm P}}}}
\def\IH{{\hbox{{\rm I}\kern-.4em\hbox{\rm H}}}}
\def\ID{{\hbox{{\rm I}\kern-.2em\hbox{\rm D}}}}

\def\be{\begin{equation}}
\def\ee{\end{equation}}
\def\ba{\begin{eqnarray}}
\def\ea{\end{eqnarray}}

\def\half{\frac{1}{2}}

\newcommand{\inv}[1]{\frac{1}{#1}}

\def\ra{\rightarrow}

\newcommand{\ud}{\mbox{${\mathrm{d}}$}}

\def\dell{\partial}

\newcommand{\bra}[1]{\mbox{$\langle #1 |$}}
\newcommand{\ket}[1]{\mbox{$| #1 \rangle$}}

\def\Tr{{\rm tr}\,}

\def\nn{\nonumber}
\def\ea{{\it et al}. }

\def\DeB{\overline{\textrm{D8}}}

\newcommand{\Mkk}{M_{\rm KK}}
\newcommand{\wt}{\widetilde}
\newcommand{\wh}{\widehat}

\begin{document}

\begin{titlepage}

\vspace{0.5in}

\begin{center}
{\large \bf Electromagnetic Baryon Form Factors from Holographic QCD} \\
\vspace{10mm}
  Keun-Young Kim and Ismail Zahed\\
\vspace{5mm}
{\it Department of Physics and Astronomy, SUNY Stony-Brook, NY 11794}\\
\end{center}
\begin{abstract}
In the holographic model of QCD suggested by Sakai and Sugimoto,
baryons are chiral solitons sourced by D4 instantons in bulk of
size $1/\sqrt{\lambda}$ with $\lambda=g^2N_c$. We  quantize the
D4 instanton semiclassically using $\hbar=1/(N_c\lambda)$ and
non-rigid constraints on the vector mesons. The holographic
baryon is a small chiral bag in the holographic direction with a
Cheshire cat smile. The vector-baryon interactions occur at the
core boundary of the instanton in $D4$. They are strong and of
order $1/\sqrt{\hbar}$. To order $\hbar^0$ the electromagnetic
current is entirely encoded on the core boundary and vector-meson
dominated. To this order, the electromagnetic charge radius is of
order $\lambda^0$. The meson contribution to the baryon magnetic
moments sums identically to the core contribution. The proton and
neutron magnetic moment are tied by a model independent relation
similar to the one observed in the Skyrme model.
\end{abstract}

\end{titlepage}

\renewcommand{\thefootnote}{\arabic{footnote}}
\setcounter{footnote}{0}



\section{Introduction}

Holographic QCD has provided an insightful look to a number of
issues in baryonic physics at strong coupling $\lambda=g^2N_c$
and large number of colors $N_c$~\cite{BRODSKY,RHO, Sakai3, Hata,
KSZ, NAWA, Others}. In particular, in \cite{BRODSKY,RHO} baryons
are constructed from a five-dimensional Shrodinger-like equation
whereby the 5th dimension generates mass-like anomalous
dimensions through pertinent boundary conditions. A number of
baryonic properties have followed ranging from baryonic spectra
to form factors \cite{BRODSKY,RHO}.

At large $N_c$ baryons are chiral solitons in QCD. A particularly
interesting framework for discussing this scenario is the
D8-$\DeB$ chiral holographic model recently suggested by Sakai and
Sugimoto~\cite{ Sakai3,Sakai1,Sakai2} (herethrough hQCD). In hQCD
D4 static instantons in bulk source the chiral solitons or
Skyrmions on the boundary. The instantons have a size of order
$1/\sqrt{\lambda}$ and a mass of order $N_c\lambda$ in units of
$M_{KK}$, the Kaluza-Klein scale~\cite{Sakai3}. The static
Skyrmion is just the instanton holonomy in the Z-direction, with
a larger size of order $\lambda^0$~\cite{SEITZ}.

In this paper we would like to elaborate further on the precedent
observation by explicitly constructing the pertinent
electromagnetic current for a holographic soliton following from
the exact D4 instanton in bulk in a semiclassical expansion with
$\hbar=1/\l N_c$. The vector mesons are quantized using non-rigid
constraints to preserve causality. The electromagnetic current is
boundary valued as expected from the solitonic nature of the
baryon as well as the holographic principle. To order $\hbar^0$
the current is entirely vector meson dominated in overall
agreement with the effective analysis in~\cite{RHO}. Our
semiclassical analysis provides a systematic framework for
analyzing the baryons in holographic QCD. It also clarifies  a
recent analysis~\cite{Hata}.

In section 2 we briefly go over the soliton-instanton
configuration of the Yang-Mills-Chern-Simons effective
theory of the Sakai-Sugimoto model in 5 dimensions,
including some generic symmetries of the instanton
configuration. In section 3 we detail our semiclassical
analysis to order $\hbar^0$. In section 4 we derive the
baryon current also to order $\hbar^0$, and show that
it is vector meson dominated. In section 5 we derive
the electromagnetic form factor and show that the minimum
and magnetic vector couplings are tied by the solitonic
nature of the baryon to order $\hbar^0$. In section 6
the electromagetic charge and radius are worked out.
While the instanton in bulk carries a size of order
$1/\sqrt{\l}$, its holographic image the baryon carries a
size of order $\l^0$ thanks to the trailing vector
mesons. The baryon magnetic moments are given in section 7.
Our conclusions are in section 8. Some useful details can
be found in the Appendices.

\section{5D YM-CS Model}

\subsection{Action}

In this section we review the action and its soliton solution
obtained in~\cite{Sakai3}. We start with the
Yang-Mills-Chern-Simons(YM-CS) theory in a 5D curved background,
which has been derived as an effective theory of
Sakai-Sugimoto(SS) model~\cite{Sakai1,Sakai3}. The 5D Yang-Mills
action is the leading terms in the $1/\l$ expansion of the DBI
action of the D8 branes after integrating out the $S^4$. The 5D
Chern-Simons action is obtained from the Chern-Simons action of
the D8 branes by integrating $F_4$ RR flux over the $S^4$, which
is nothing but $N_C$. The action reads~\cite{Sakai1,Sakai3}
\begin{eqnarray}
&&S = S_{YM} + S_{CS}\ ,  \label{YM-CS}\\
&&S_{YM} = - \k \int d^4x dZ \ \Tr \left[\half K^{-1/3}
\calf_{\m\n}^2 + \Mkk^2 K
\calf_{\m Z}^2 \right] \ , \label{YM} \\
&&S_{CS} = \frac{N_c}{24\pi^2}\int_{M^4 \times R}
\w_5^{U(N_f)}(\cala) \ , \label{CS}
\label{REDUCED}
\end{eqnarray}
where $\m,\n = 0,1,2,3$ are 4D indices and the fifth(internal)
coordinate $Z$ is dimensionless.  There are three things which are
inherited by the holographic dual gravity theory: $\Mkk, \k,$ and
$K$. $\Mkk$ is the Kaluza-Klein scale and we will set $\Mkk = 1$
as our unit. $\k$ and $K$ are defined as
\begin{eqnarray}
\k = {\l N_c} \inv {216 \pi^3} \equiv \l N_c a  \ , \qquad K = 1
+ Z^2 \ .
\end{eqnarray}
$\cala$ is the 5D $U(N_f)$ 1-form gauge field and $\calf_{\m\n}$
and $\calf_{\m Z} $ are the components of the 2-form field
strength $\calf = \ud \cala -i \cala \wedge \cala$.
$\w_5^{U(N_f)}(\cala)$ is the Chern-Simons 5-form for the $U(N_f)$
gauge field:
\begin{eqnarray}
  \w_5^{U(N_f)}(\cala) = \Tr \left( \cala \calf^2 + \frac{i}{2} \cala^3 \calf - \inv{10} \cala^5
  \right)\ ,
\end{eqnarray}
Since $\cala$ is $U(N_f)$ valued, it may be decomposed into an
$SU(N_f)$ part($A$) and a $U(1)$ part($\widehat{A}$),
\begin{eqnarray}
\cala =  A + \frac{1}{\sqrt{2 N_f}}  \widehat{A} \ , \qquad \calf
= F +  \frac{1}{\sqrt{2 N_f}}  \widehat{F} \ ,
\end{eqnarray}
where $A \equiv A^a T^a , F \equiv F^a T^a$ and the $SU(N_f)$
generators $T^a$ are normalized as
\begin{eqnarray}
\Tr (T^a T^b) = \half\d^{ab} \ .
\end{eqnarray}

For $N_f=2$ the action (\ref{YM}) and (\ref{CS}) are reduced to
\begin{eqnarray}
S_{YM} &=& -\k \int d^4x dZ \ \Tr \left[\half K^{-1/3} F_{\m\n}^2
+ K F_{\m Z}^2 \right] \nn \\
&& -\frac{\k}{2} \int d^4x dZ \ \left[\half {K}^{-1/3}
\widehat{F}_{\m\n}^2 + {K} \widehat{F}_{\m Z}^2 \right]  \ ,
\label{YM1}
\end{eqnarray}
\begin{eqnarray}
S_{CS} &=& \frac{N_c}{24\pi^2}\int \left[
\frac{3}{2}\widehat{A}\Tr F^2 +
\frac{1}{4}\widehat{A}\widehat{F}^2 + \frac{1}{2}\ \ud \left\{
\widehat{A}\ \Tr \left(2FA+\frac{i}{2}A^3 \right) \right\}
\right] \label{CS1}  \\
&=&  \frac{N_c}{24\pi^2}\e_{MNPQ} \int d^4x dZ \left[
\frac{3}{8}\wh{A}_0 \Tr (F_{MN}F_{PQ}) - \frac{3}{2}\wh{A}_M \Tr
(\dell_0 A_N F_{PQ})
\right. \nn \\
&& + \frac{3}{4}\wh{F}_MN \Tr (A_0 F_{PQ}) + \frac{1}{16} \wh{A}_0
\wh{F}_{MN}\wh{F}_{PQ} - \frac{1}{4}\wh{A}_M
\wh{F}_{0N}\wh{F}_{PQ}
\nn \\
&& \left. + \frac{3}{2}\dell_N(\wh{A}_M \Tr A_0F_{PQ}) \right]  +
 \frac{N_c}{48\pi^2} \int  \ud \left\{ \widehat{A}\ \Tr
\left(2FA+\frac{i}{2}A^3 \right) \right\} \ , \label{CS2}
\end{eqnarray}
where the $SU(2)$ and $U(1)$ parts are completely desentangled in the
Yang-Mills action. The $ \w_5^{SU(2)}(A)$ vanishes in the CS
action. The action (\ref{YM1}) and (\ref{CS1}) yield the 10
coupled equations of motion, of which the D4 instanton  is
a solution with topological charge 1. The coupled equations
are specifically given by
\begin{eqnarray}
&&\d A_0 \ra \ \k \left\{ {D}^\m \left( K^{-1/3} {F}_{\m 0}
\right) + {D}^Z \left( K {F}_{Z 0} \right) \right\} -
\frac{N_c}{64\pi^2}\e_{MNPQ}( \wh{F}_{MN} F_{PQ}) = 0 \ ,
\label{EOM:Full:1}
 \\
&&\d A_i \ra \ \k \left\{ {D}^\m \left( K^{-1/3} {F}_{\m i}
\right) + {D}^Z \left( K {F}_{Z i} \right) \right\} -
\frac{N_c}{64\pi^2}\e_{iNPQ}( \wh{F}_{N0} F_{PQ} + \wh{F}_{PQ}
F_{N0}) = 0 \ , \label{EOM:Full:2}
\\
&&\d A_Z \ra\ \k \left\{ {D}^\m \left( K {F}_{\m Z} \right)
\right\} - \frac{N_c}{64\pi^2}\e_{ZNPQ}( \wh{F}_{N0} F_{PQ} +
\wh{F}_{PQ}
F_{N0}) = 0 \ ,   \label{EOM:Full:3} \\
&&\d \widehat{A}_0 \ra\ \k \left\{ {\dell}^\m \left(K^{-1/3}
{\widehat{F}}_{\m 0} \right) + \dell^Z \left( K {\widehat{F}}_{Z
0} \right) \right\}
\nn \\
&& \qquad \qquad \qquad \qquad \qquad \qquad  -
\frac{N_c}{64\pi^2}\e_{MNPQ}\Big( \Tr (F_{MN} F_{PQ}) + \half
\widehat{F}_{MN} \widehat{F}_{PQ}  \Big) = 0 \ ,
\label{EOM:Full:4}
\\
&&\d \widehat{A}_i \ra\ \k \left\{ \dell^\m \left( K^{-1/3}
{\widehat{F}}_{\m i} \right) + \dell^Z \left( K {\widehat{F}}_{Z
i} \right) \right\}
\nn \\
&&  \qquad \qquad \qquad \qquad \qquad \qquad -
\frac{N_c}{16\pi^2}\e_{iNPQ}\Big( \Tr(F_{N0} F_{PQ}) + \half
\wh{F}_{N0} \wh{F}_{PQ}\Big) = 0 \ , \label{EOM:Full:5}
\\
&&\d \widehat{A}_Z \ra\ \k \left\{ \dell^\m \left( K
{\widehat{F}}_{\m Z} \right) \right\} -
\frac{N_c}{16\pi^2}\e_{ZNPQ}\Big( \Tr(F_{N0} F_{PQ}) + \half
\wh{F}_{N0} \wh{F}_{PQ}\Big) = 0 \ . \label{EOM:Full:6}
\end{eqnarray}
We note that $\delta A_0$ and $\delta\widehat{A}_0$ are constraint
type equations or Gauss laws.

\subsection{The Instanton Solution}

The exact static $O(4)$ solution in $x^M$ space in the large $\l$
limit is not known. Some generic properties of this solution can
be derived for large $\l$ whatever the curvature. Indeed, since
$\k \sim \l$, the instanton solution with unit topological charge
that solves (\ref{EOM:Full:1}-\ref{EOM:Full:6}) follows from the
YM part only in leading order. It has zero size at infinite $\l$.
At finite $\l$ the instanton size is of order $1/\sqrt{\l}$. The
reason is that while the CS contribution of order $\l^0$ is
repulsive and wants the instanton to inflate, the warping in the
Z-direction of order $\l^0$ is attractive and wants the instanton
to deflate in the Z-direction~\cite{RHO,Sakai3}.

For some insights to the warped instanton configuration at
large $\l$ we follow~\cite{Sakai3} and rescale the coordinates and
the $U(2)$ gauge fields $\cala$ as
\begin{eqnarray}
&&x^M = \l^{-1/2}\widetilde{x}^M \ , \quad x^0 = \widetilde{x}^0 \ , \nn \\
&&\cala_M = \l^{1/2} \widetilde{\cala}_M \ , \quad \cala_0 = \widetilde{\cala}_0 \ , \nn \\
&& \calf_{MN} = \l \widetilde{\calf}_{MN} \ , \quad \calf_{0M} =
\l^{1/2} \widetilde{\calf}_{0M} \ ,
\label{Rescale}
\end{eqnarray}
where $M,N = 1,2,3,Z$ and $x^Z \equiv Z$. The variables with
tilde are of order of $\l^0$.
The equations of motions of order $\l$ are
\begin{eqnarray}
&& \widetilde{D}^N \widetilde{F}_{MN}  = 0 \ ,  \label{lambdaorder1} \\
&& \widetilde{\dell}^N \wh{\widetilde{F}}_{MN}  = 0 \ ,
\label{lambdaorder2}
\end{eqnarray}
which yield $\wh{\wt{\mathbb{A}}}_M = 0$
\footnote{For clarity we summarize our convention here. Greek
indices $\{\mu,\nu\}=0,1,2,3,4$, capital latin indices
$\{M,N,P,Q\} = 1,2,3,4$, and small latin indices $\{i,j,k\} =
1,2,3$. The fifth coordinate $Z$ has index 4. i.e. $4 \equiv Z$.
The gauge field and field strength with hat are $U(1)$ valued and
without hat they are $SU(2)$ valued.  All variables with tilde
are of order of $\l^0$ and without tilde they behave as
(\ref{Rescale}). We denote the classical field by the boldface.
(e.g $\mathbb{A}$). $A$ and $F$ are understood as form without
component indices.}
for the $U(1)$ part and the BPST instanton solution for the $SU(2)$
part:
\begin{eqnarray}
\widetilde{\mathbb{A}}_M = \eta_{iMN}\frac{\sigma_i}{2}\frac{2
\widetilde{x}_N}{\widetilde{\xi}^2 + \widetilde{\rho}^2} \ ,
\qquad \widetilde{\mathbb{F}}_{MN} =
\eta_{iMN}\frac{\sigma_i}{2}\frac{-4\widetilde{\rho}^2}{(\widetilde{\xi}^2
+ \widetilde{\rho}^2)^2} \ . \label{ADHM1}
\end{eqnarray}
The instanton is located at the origin so $\widetilde{\xi}
\equiv \sqrt{\vec{\widetilde{x}}^2 + \widetilde{Z}^2}$.
$\eta_{iMN}$ is t'Hooft symbol defined as $\eta_{ijk} \equiv
\e_{ijk}$, and $\eta_{iMZ} = \d_{iM}$.
At this order $\wt{A}_0$ and $\wh{\wt{A}}_0$ are not determined
and there is no restriction on the size of the BPST instanton.

The equations of motion to order $\l^0$ are
\begin{eqnarray}
&& \widetilde{D}^2_M
\widetilde{A}_{0} = 0 \ ,  \label{sub1} \\
&&
\widetilde{\dell}^2_M\wh{\widetilde{A}}_{0}-\frac{1}{64\pi^2a}\e_{MNPQ}
\Tr ( \widetilde{F}_{MN}\widetilde{F}_{PQ})  = 0 \ ,   \label{sub2} \\
&& \frac{2}{3}\widetilde{Z}^2\widetilde{D}_j\widetilde{F}_{ij} + 2
\widetilde{Z}^2 \widetilde{D}_Z\widetilde{F}_{iZ} +
2\widetilde{D}_0\widetilde{F}_{0i} - \frac{1}{8\pi^2 a }
\e_{ijkZ} \wh{\widetilde{A}}_0
(\widetilde{D}_j\widetilde{F}_{kZ}) = 0 \ , \label{sub3}
\\
&& - 2 \widetilde{Z}^2 \widetilde{D}_i\widetilde{F}_{iZ} +
2\widetilde{D}_0\widetilde{F}_{0Z} - \frac{1}{8\pi^2 a } \e_{ijkZ}
\wh{\widetilde{A}}_0 (\widetilde{D}_k\widetilde{F}_{ij}) = 0 \ ,
\label{sub4}
\end{eqnarray}
Gauss law (\ref{sub1}) and (\ref{sub2}) fix
$\widetilde{\mathbb{A}}_{0}$ and
$\widehat{\widetilde{\mathbb{A}}}_{0}$ as
\begin{eqnarray}
\widetilde{\mathbb{A}}_{0} = 0 \ , \qquad
\widehat{\widetilde{\mathbb{A}}}_{0} = -\frac{1}{8 \pi^2 a
}\frac{2\widetilde{\rho}^2 +
\widetilde{\xi}^2}{(\widetilde{\rho}^2 + \widetilde{\xi}^2)^2} \
. \label{A0sol}
\end{eqnarray}
To this order, the leading BPST solution together with
(\ref{A0sol}) solve (\ref{sub3}) and (\ref{sub4}) for fixed size
$\tilde\rho$ of order $\lambda^0$. Equivalently, this size
follows from the the minimum of the energy to order
$1/\l$~\cite{Sakai3}.
%
%
For completeness, we note in this section  that in terms of the
unrescaled variables the instanton gauge fields are
\begin{eqnarray}
&& \mathbb{A}_0 = 0 \ , \qquad \qquad \qquad\qquad\quad
\mathbb{A}_M = \eta_{iMN}\frac{\sigma_i}{2}\frac{2
x_N}{\xi^2 + \rho^2} \ , \label{ADHM2}  \\
&&\widehat{\mathbb{A}}_0 = -\frac{1}{8 \pi^2 a \l}\frac{2\rho^2 +
\xi^2}{(\rho^2 + \xi^2)^2} \ , \qquad  \wh{\mathbb{A}}_M = 0
\label{A02} \ ,
\end{eqnarray}
and the nonvanishing field strengths are
\begin{eqnarray}
\mathbb{F}_{MN} =
\eta_{iMN}\frac{\sigma_i}{2}\frac{-4\rho^2}{(\xi^2 + \rho^2)^2} \
, \qquad  \widehat{\mathbb{F}}_{M0} = \frac{x^M (3\rho^2 +
\xi^2)}{4 \pi^2 a \l (\rho^2 + \xi^2)^3} \ ,  \label{ADHM3}
\end{eqnarray}
with the size $\rho=\tilde{\rho}/\sqrt{\l}$,
\begin{eqnarray}
\rho^2 =  \frac{1}{8\pi^2 a \l} \sqrt{\frac{6}{5}} \ .
\end{eqnarray}
We note that near the origin $\xi\sim 0$, the field strengths are
large with $\mathbb{F}\sim 1/\rho^2\sim \l$ and
$\widehat{\mathbb{F}}\sim 1/(\l \rho^4)\sim {\l}$. In a way the
reduced DBI action (\ref{REDUCED}) is not justified for such
field strengths since higher powers of the field strength
contribute. For our semiclassical analysis below this does not
really matter, since an exact solution of the instanton problem
with the full DBI action will not affect the generic nature of
most results below.

For large $Z$ and finite $\lambda$, the warped instanton
configuration is not known. While we do not need it for the
semiclassical analysis we will detail below, some generic
properties can be inferred. Indeed, the small-Z BPST
configuration above has maximal spherical symmetry. That is that
an isospin rotation is equivalent to (minus) a space rotation, a
feature that is immediately checked through
\begin{eqnarray}
(\mathbb{R}\mathbb{A})_Z = \mathbb{A}_Z(\mathbb{R} \vec{x}) \ ,
\qquad (\mathbb{R}^{ab}\mathbb{A}^{b})_i =
\mathbb{R}^T_{ij}\mathbb{A}_j^a(\mathbb{R}\vec{x})\,\,,
\end{eqnarray}
with $\mathbb{R}$ a rigid $SO(3)$ rotation.
When semiclassically quantized, the instanton-baryon
configuration yields a tower of states with isospin matching
minus the spin. This is expected, since the holographic instanton
is a Skyrmion on the boundary with hedgehog symmetry. These
symmetries can be used to construct variationally the warped
instanton configuration, a point we will present elsewhere.

\section{Non-rigid SemiClassical Expansion}

In this section we assume that the instanton configuration $\mathbb{A}$
solves exactly the equations of motion for all $Z$ and all $\l$ and
proceed to quantize it semiclassically using $\hbar=1/\k\sim 1/\l N_c$.
For the book-keeping to work we count $\rho^2$ of order $\hbar^0$.
Since the holographic pion decay constant $f_\pi^2\sim \k$, this
is effectively the analogue of the semiclassical $1/N_c$ expansion
of the boundary Skyrmion, albeit at strong $\l$ coupling.

We now note that $\mathbb{A}$ exhibits exact flavor,
translational and rotational zero modes as well as soft or
quasi-zero modes in the size $\rho$ and conformal direction $Z$.
We will use collective coordinates to quantize them in general.
While for the electromagnetic analysis below we focus on the
isorotations (minus the spatial rotations) only, we will discuss
in this section the semiclassical anlysis in general.

\begin{figure}[]
  \begin{center}
    \includegraphics[width=11cm]{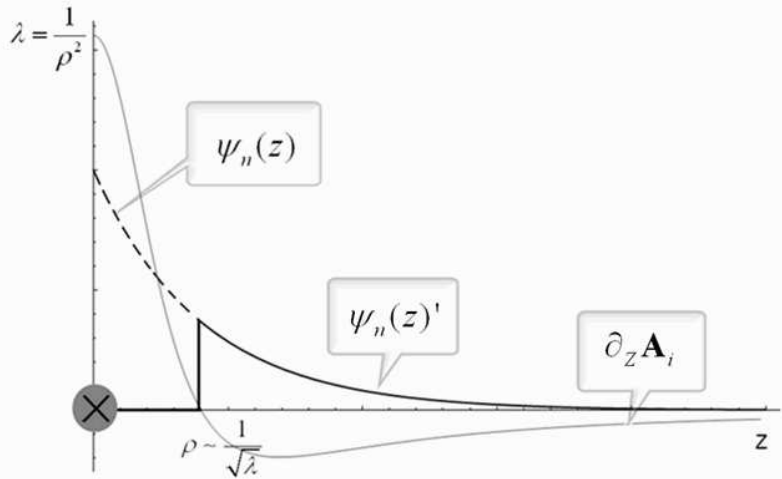}
  \caption{The Z-mode in the non-rigid gauge vs $\dell_Z \mathbb{A}_i$.}
  \label{Fig:fig1}
  \end{center}
\end{figure}

Generically, we have in the body fixed frame
\begin{eqnarray}
A_M(t,x,Z) = \mathbb{R}(t) \left(\mathbb{A}_M(x-X_0(t),Z-Z_0(t))
+ C_M(t,x-X_0(t),Z-Z_0(t))\right) \ , \label{SEM1}
\end{eqnarray}
with $\rho=\rho(t)$. The classical part transforms
inhomogeneously under flavor gauge transformation, while the
quantum part transforms homogeneously. The fluctuations $C$ are
quantum and of order $\sqrt{\hbar}$ (see below). The isoration
$\mathbb{R}$ is an $SO(3)$ matrix which is the adjoint
representation of the $SU(2)$ flavor group. Its generators are
real $(G^B)^{ab}=\epsilon^{aBb}$. To order $\hbar^0$ the
constrained field $\widehat{\mathbb{A}}_0$ remains unchanged,
while the constrained field ${\mathbb{A}}_0=0$ shifts by a
time-dependent zero mode as detailed in Appendix A. The
collective coordinates $\mathbb{R}, X_0, Z_0, \rho$ and the
fluctuations $C$ in (\ref{SEM1}) form a redundant set. Indeed,
the true zero modes
\begin{eqnarray}
\delta^B_R\mathbb{A}_M = G^B\mathbb{A}_M \ , \qquad
\qquad\delta^i\mathbb{A}_M=\nabla^i\mathbb{A}_M \ , \label{SEM2}
\end{eqnarray}
and the quasi-zero or soft modes are
\begin{eqnarray}
\delta_Z\mathbb{A} = \partial_Z\mathbb{A} \ , \qquad
\qquad\delta_\rho\mathbb{A}_M=\partial_\rho\mathbb{A}_M \ ,
\label{SEM3}
\end{eqnarray}
modulo gauge transformations. All the analysis to follow will be
carried to order $\hbar^0$. To avoid double counting, we need to orthogonalize
(gauge fix) the vector fluctuations $C$ from
(\ref{SEM2}-\ref{SEM3}) though pertinent constraints. In the {\it
rigid quantization}, the exact zero modes are removed from the
spectrum of $C$. For example, the isorotations are removed by the
constraint
\begin{eqnarray}
\int d\xi\, C\,G^B\mathbb{A}_M \ , \label{SEM4}
\end{eqnarray}
and similarly for the translations. For an application to chiral
baryons of this method we refer to~\cite{ADAMI}.
This constaint violates causality
as the fluctuation orthogonalizes instantaneously to an infinitesimal
isorotation throughout the instanton body. A causal semiclassical
quantization scheme has been discussed in~\cite{VERSCHELDE}. Here, it
means that for instance (\ref{SEM4}) should only be enforced at the
location of the instanton, i.e.
\begin{eqnarray}
\int_{x=Z=0} d{\hat{\xi}} C\,G^B\mathbb{A}_M\,\,.
\label{SEM5}
\end{eqnarray}
For $Z$ and $\rho$ the non-rigid constraints are more natural to
implement since these modes are only soft near the origin at large
$\l$. The vector fluctuations at the origin linearize through the modes
\begin{eqnarray}
d^2\psi_n/dZ^2= \l_n\psi_n \ , \label{SEM6}
\end{eqnarray}
with $\psi_n(Z)\sim e^{-\sqrt{\l_n}Z}$. In the spin-isospin 1
channel they are easily confused with $\partial_Z\mathbb{A}_i$
near the origin as we show in Fig.~\ref{Fig:fig1}.
Using the non-rigid constraint,  the double counting is removed
by removing the origin from the vector mode functions
\begin{eqnarray}
\psi_n'(Z)=\theta(|Z|-Z_c)\psi_n(Z) \ , \label{SEM7}
\end{eqnarray}
with $Z_C\sim\rho\sim 1/\sqrt{\l}$ which becomes the origin for
large $\l$. In the non-rigid semiclassical framework, the baryon
at small $\xi<|Z_C|$ is described by an instanton located at the
origin of $R^4$ and rattling in the vicinity of $Z_C$. At large
$\xi>|Z_C|$, the rattling instanton sources the vector meson
fields described by a semi-classical expansion with non-rigid
Dirac constraints. Changes in $Z_c$ (the core boundary) are
reabsorbed by a residual gauge transformation on the core
instanton. This is a holographic realization of the Cheshire cat
principle~\cite{CAT} where $Z_c$ plays the role of the Cheshire
cat smile.

For simplicity, and throughout
the semiclassical expansion we will ignore the translational zero
modes. Also for simplicity, the expansion will not rely on the
Dirac constraint of the isorotations since to order $\hbar^0$
their contribution does not arise in the electromagnetic current.
A similar observation was made in the Skyrme model where
pion-baryon couplings are found to be leading and
time-like~\cite{SCHNITZER,VERSCHELDE}. At order $\hbar$ and
higher the Dirac constraints matter. The constraint on the Z-mode
is implemented by $Z_C$ throughout.

To order $\hbar^0$ the semiclassical expansion will be carried
out covariantly in the action formalism, whereby Gauss law is
unfolded for both $\widehat{\mathbb A}_0$ and ${\mathbb A}_0$
as detailed in Appendix A. To this order there is no difference
between the canonical Hamiltonian formalism, with the advantage
of manifest covariance for the derived flavor currents.

Having said this, we now use the gauge field
decomposition presented in~\cite{Sakai2} for the
non-rigid semiclassical expansion and refer to this
work for further references. Specifically,
\begin{eqnarray} \label{GF1}
A_\m &=& \mathbb{A}_\m + C_\m \ , \qquad   C_\m \equiv
v_\m^n\psi_{2n-1} + a_\m^n \psi_{2n} + \calv_\m  + \scra_\m \psi_0
 \ , \\  \label{GF2} A_Z &=& \mathbb{A}_Z  + C_Z \ , \qquad
C_Z \equiv - i\Pi \phi_0 \ ,
\\  \widehat{A}_\m &=& \widehat{\mathbb{A}}_\m + \widehat{C}_\m
\ , \qquad \widehat{C}_\m \equiv \widehat{v}_\m^n\psi_{2n-1} +
\widehat{a}_\m^n \psi_{2n} + \widehat{\calv}_\m  +
\widehat{\scra}_\m \psi_0 \ ,\\  \label{GF3} \widehat{A}_Z &=&
\widehat{\mathbb{A}}_Z  + \widehat{C}_Z \ , \qquad \widehat{C}_Z
\equiv - i\widehat{\Pi} \phi_0 \ ,\label{GF4}
\end{eqnarray}
where $\{\mathbb{A} , \widehat{ \mathbb{A}}\}$ refer to the
instanton configuration and $\{C , \widehat{C}\}$ to the vector
meson fluctuations. The $\mathbb{R}$ rotation is subsumed. $\{v^n,
\widehat{v}^n\}$, $\{a^n,\widehat{a}^n\}$, and $\{\Pi ,
\widehat{\Pi} \}$ are the vector mesons, the axial vector mesons
and the pions respectively. $\{ \calv,\widehat{ \calv} \}$ is the
vector source and  $\{ \scra,\widehat{\scra} \}$ is the axial
vector source. Theses meson and source fields are all functions
of $x^\m$. They are attached to the mode functions $\{\psi,
\phi\}$ in bulk which are functions of $Z$ as expounded
in~\cite{Sakai2}:
\begin{eqnarray}
&& -K^{1/3} \dell_Z (K \dell_Z \psi_n) = m^2_{v^n} \psi_n \ ,
\qquad
\k\int dZ K^{-1/3}\psi_n\psi_m = \d_{nm} \ , \label{EigenEqn} \\
&& \a_{v^n} \equiv \k\int dZ K^{-1/3}\psi_{2n-1}\ , \qquad
\a_{v^n} \equiv \k\int dZ K^{-1/3}\psi_{2n}\psi_0 \ ,  \\
&& \psi_0 \equiv \frac{2}{\pi}\arctan Z \ , \qquad \phi_0 \equiv
\frac{1}{\sqrt{\pi\k}K} \ .
\end{eqnarray}

With the gauge field (\ref{GF1})-(\ref{GF4}) the $ SU(2)$ YM
action reads
\begin{eqnarray}
S_{YM} &=&  - \frac{\k}{2}  \int d^4x dZ   \left[ \dell^Z
\left(2K \widehat{\mathbb{F}}_{Z\m}\widehat{C}^\m\right) \right. \nn \\
&& \qquad \qquad\qquad \left.   +  \half K^{-1/3} \Big( \dell_\m
\widehat{C}_\n - \dell_\n \widehat{C}_\m \Big)^2 + K  \Big(
\dell_Z \widehat{C}_\m - \dell_\m \widehat{C}_Z \Big)^2 \right] \
, \\ \label{U1B}
& &-\k  \int d^4x dZ  \Tr  \left[ \dell^Z
\left(2K \mathbb{F}_{Z\n}C^\n \right) \right.  \nn \\
&& \qquad \qquad\qquad  +  \half K^{-1/3}
\left\{2\mathbb{F}_{\m\n} [C^\m, C^\n] + \Big( \mathbb{D}_\m C_\n
- \mathbb{D}_\n C_\m -i [C_\m, C_\n]\Big)^2
\right\} \nn \\
&& \qquad \qquad\qquad  \left. + K \left\{2\mathbb{F}_{Z\m} [C^Z,
C^\m] + \Big( \mathbb{D}_Z C_\m - \mathbb{D}_\m C_Z -i [C_Z,
C_\m]\Big)^2 \right\} \right] \label{SU2} \ ,
\end{eqnarray}
where $ \mathbb{D}_\a $ is the covariant derivative with the
slowly rotating instanton in flavor space: $\mathbb{D}_\a * =
\dell_\a -i[\mathbb{A}_\a , *]$. We dropped the leading and pure
instanton part for convenience. Some details regarding the
expansion including the CS part are briefly given in Appendix B.
All the linear terms to $\{C , \widehat{C}\}$ except the {\it
boundary terms} in the YM and CS action vanish due to the
equations of motion. There is no coupling in bulk between the
vector mesons and the instanton configuration except through
boundary terms and/or time derivatives. This is the hallmark of
solitons. While it apparently looks different from the effective
and holographic description presented in~\cite{RHO} as couplings
are involved in bulk, we will show below that the results are
indeed similar for the electromagnetic form factors to order
$\hbar^0$. Below, we will explain why the similarity.

We note that without the instanton $\{\mathbb{A} , \widehat{
\mathbb{A}}\}$, the action reduces to the one in~\cite{Sakai2},
where the vector meson dominance (VMD) of the pion form factor
follows from the field redefinitions
\begin{eqnarray}
&& v^n \ra v_{\mathrm{new}}^n = v^n + \a_{v^n}\calv +
\frac{b_{v^n\pi\pi}}{2f_\pi^2}[\Pi, \ud \Pi] \ , \\
&& b_{v^n\pi\pi} \equiv \k \int dZ K^{-1/3}(1-\psi_0^2) \ ,
\end{eqnarray}
with $f_\pi$ the pion decay constant. This redefinition
yields a direct vector-photon coupling $v^n-\calv$
\begin{eqnarray}
m_{{v}^n}^2 \left({v}^n_{\mathrm{new}}    -\a_{v^n} {\calv}
\right)^2 \ ,
\end{eqnarray}
while removing all pion-photon couplings $\Pi-\calv$ through
various sum rules. Holographic QCD obeys the strictures of VMD in
the meson sector. This point will carry semiclassically to the
baryon sector as we detail below.

Substituting the 5D fields for the 4D fields with mode functions we
have to order $\hbar^0$
\begin{eqnarray}
&& S_{\mathrm{eff}} = \nn \\
&& \ - \sum_{n=1}^{\infty} \int d^4x \left[ \left[\k K
\widehat{\mathbb{F}}^{Z\m} \left( (\widehat{v}_\m^n -
\a_{v^n}\widehat{\calv}_\m) \psi_{2n-1} + (\widehat{a}_\m^n -
\a_{a^n}\widehat{\scra}_\m)\psi_{2n} + \widehat{\calv}_\m  +
\widehat{\scra}_\m \psi_0  \right) \right]_{Z=B} \right. \nn \\
&& \qquad \qquad \quad + \frac{1}{4} \Big( \dell_\m
\widehat{v}^n_\n - \dell_\n \widehat{v}^n_\m \Big)^2 +
\frac{1}{2} m_{{v}^n}^2 (\widehat{v}^n_\m  -\a_{v^n} \widehat{\calv}_\m   )^2 \nn \\
&&\qquad \qquad \quad \left.  + \frac{1}{4} \Big( \dell_\m
\widehat{a}^n_\n - \dell_\n \widehat{a}^n_\m \Big)^2 +
\frac{1}{2} m_{{a}^n}^2 \left(\widehat{a}^n_\m  -\a_{a^n}
\widehat{\scra}_\m     \right)^2 \right]  \ , \label{U1B} \\
&&- \  \sum_{n=1}^{\infty} 2 \Tr  \int d^4x \left[ \left[ \k K
\mathbb{F}^{Z\n} \left( (v_\m^n  - \a_{v^n}\calv_\m) \psi_{2n-1}
+ (a_\m^n  - \a_{a^n}\scra_\m)  \psi_{2n} + \calv_\m + \scra_\m
\psi_0
 \right) \right]_{Z=B} \right.   \nn \\
&& \qquad \qquad \quad  + \frac{1}{4} \Big( \dell_\m {v}^n_\n -
\dell_\n {v}^n_\m \Big)^2 + \frac{1}{2}
m_{{v}^n}^2 \left({v}^n_\m    -\a_{v^n} {\calv}_\m   \right)^2 \nn \\
&& \qquad \qquad \quad \left. + \frac{1}{4} \Big( \dell_\m
{a}^n_\n - \dell_\n {a}^n_\m \Big)^2 + \frac{1}{2}
 m_{{a}^n}^2 \left({a}^n_\m   -\a_{a^n} {\scra}_\m  \right)^2 \right]
\label{SU2} \ ,
\end{eqnarray}
where all meson fields are the redefined fields,
$v^n_{\mathrm{new}}$ and $a^n_{\mathrm{new}}$, but we drop the
subscript for simplicity. $[\cdots]_{Z=B}$ will be evaluated at
the boundary $Z=B$, which is collectively denoted by $\{ \pm
\infty, \pm Z_c \}$.  We retained only the terms relevant to the
baryon form factor. For the complete expansion of the meson
fluctuation part we refer to~\cite{Sakai2}.

\section{Baryon current}

\begin{figure}[]
  \begin{center}
    \includegraphics[width=11cm]{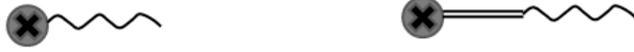}
  \caption{Left: Direct coupling, Right: Vector meson mediated coupling(VMD) }
  \label{Fig:FF}
  \end{center}
\end{figure}

Now, consider the effective action for the $U(1)_V$ source to
order $\hbar^0$
\begin{eqnarray}
&& S_{\mathrm{eff}}[\widehat{\calv}_\m] =  \sum_{n=1}^{\infty}
\int d^4x \left[ -\frac{1}{4} \Big( \dell_\m \widehat{v}^n_\n -
\dell_\n \widehat{v}^n_\m \Big)^2 -\frac{1}{2} m_{\hat{v}^n}^2
(\widehat{v}^n_\m)^2
 \right. \nn \\
&& \qquad \qquad  \qquad \qquad \quad  -  \k K
\widehat{\mathbb{F}}^{Z \m} \widehat{\calv}_\m
(1-\a_{v^n}\psi_{2n-1})
\Big|_{Z=B}  \nn \\
&& \qquad \qquad  \qquad \qquad \quad  + \
 a_{{v}^n} m_{{v}^n}^2 \widehat{v}^n_\m \widehat{\calv}^\m
- \k K \widehat{\mathbb{F}}^{Z \m} \widehat{v}_\m^n\psi_{2n-1}
\Big|_{Z=B}  \Big] \label{U1source}
 \ ,
\end{eqnarray}
The first line is the free action of the massive vector meson
which gives the meson propagator
\begin{eqnarray}
\Delta_{\m\n}^{mn}(x) = \int \frac{d^4 p}{(2\pi)^4} e^{-ipx}
\left[ \frac{- g_{\m\n} - p_\m p_\n / m_{v^n}^2}{p^2 + m_{v^n}^2}
\d^{mn}\right] \ ,
\end{eqnarray}
in our convention. The rest are the coupling terms between the
source and the instanton: the second line is the direct coupling
(Fig.\ref{Fig:FF}(a)) and the last line corresponds to the
coupling mediated by the $U(1)$ (omega, omega', ...) vector meson
couplings (Fig.\ref{Fig:FF}(b)),
\begin{eqnarray}
\k K \widehat{\mathbb{F}}^{Z \m} \widehat{v}_\m^n\psi_{2n-1} \ ,
\end{eqnarray}
which is large and of order $1/\sqrt{\hbar}$ since
$\psi_{2n-1}\sim \sqrt{\hbar}$. When $\rho$ is set to
$1/\sqrt{\l}$ after the book-keeping noted above, the coupling
scales like $\lambda\sqrt{N_c}$, or $\sqrt{N_c}$ in the large
$N_c$ limit taken first~\footnote{The reader may object that such
strong couplings may upset the semiclassical expansion through
perturbative corrections. This is not the case when the Dirac
constraints are imposed properly as noted in~\cite{VERSCHELDE}
for the Skyrme model.}.

The direct coupling drops by the sum rule
\begin{eqnarray}
\sum_{n=1}^{\infty} \a_{v^n}\psi_{2n-1} = 1 \ , \label{sumrule1}
\end{eqnarray}
following from closure in curved space
\begin{eqnarray}
\d (Z-Z') = \sum_{n=1}^{\infty}\ \k \psi_{2n-1}(Z)
\psi_{2n-1}(Z')K^{-1/3}(Z') \ .
\label{SUM}
\end{eqnarray}
in complete analogy with VMD for the pion~\cite{Sakai2}.

The baryon current is entirely vector dominated to order $\hbar^0$
and reads
\begin{eqnarray}
J^\m_{B}(x) &=&  -\sum_{n,m}  \  m_{{v}^n}^2 a_{{v}^n} \psi_{2m-1}
\int d^4y\
 \k K  \widehat{\mathbb{F}}_{Z \n}(y,Z) \Delta^{\n \m}_{mn}(y-x) \Big|_{Z=B}
 \ .
  \label{Bcurrent2}
\end{eqnarray}
This point is in agreement with the effective holographic approach
described in~\cite{RHO}. The static baryon charge distribution is
\begin{eqnarray}
J^0_{B }(\vec{x}) = -\sum_{n}  \  \int d\vec{y}
 \ \  \frac{2 }{N_c} \k K \widehat{\mathbb{F}}_{Z 0}(\vec{y},Z) \
\Delta_n(\vec{y}-\vec{x}) \ a_{{v}^n} m_{v^n}^2 \psi_{2n-1} \
\Big|_{Z=B} \ ,
\end{eqnarray}
with
\begin{eqnarray}
\Delta_n(\vec{y}-\vec{x}) \equiv \int \frac{d\vec{p}}{(2\pi)^3}
  \frac{e^{-i\vec{p}\cdot (\vec{y}-\vec{x})}}{\vec{p}^2 +
  m_{v^n}^2} \ ,
\end{eqnarray}
and the extra factor $2/N_c$ comes from the relation between
$\wh{\calv}_\m$ and the baryon number source $\wh{B}_0(\vec{x})$:
by $\widehat{\calv}_\m = \d_{\m 0} \frac{\sqrt{2N_f}}{N_c}
\widehat{B}_0(\vec{x}) $.

\section{Electromagnetic Current and Form Factor}

In addition to the baryon current discussed above, we now need the
flavor or isospin current to construct the electromagnetic current.
For that, consider the 4 dimensional effective action for the
SU(2)-valued flavor source again to order $\hbar^0$
\begin{eqnarray}
&& S_{\mathrm{eff}}[{\calv}^a_\m] = \sum_{b=1}^{3}
\sum_{n=1}^{\infty} \int d^4x \left[ -\frac{1}{4} \Big( \dell_\m
{v}^{b,n}_\n - \dell_\n {v}^{b,n}_\m \Big)^2 -\frac{1}{2}
m_{\hat{v}^n}^2 ({v}^{b,n}_\m)^2
 \right. \nn \\
&& \qquad \qquad  \qquad \qquad \quad  + \
 a_{\hat{v}^n} m_{\hat{v}^n}^2 {v}^{a,n}_\m
{\calv}^{a,\m} - \k K {\mathbb{F}}^{b,Z \m} {v}_\m^{b,n}
\psi_{2n-1} \Big|_{Z=B}  \Big]
 \ ,
\end{eqnarray}
where the direct coupling vanishes due to the sum rule
(\ref{sumrule1}) and only the VMD part contributes through the
$SU(2)$ (rho, rho', ...) meson couplings
\begin{eqnarray}
\k K {\mathbb{F}}^{b,Z \m} {v}_\m^{b,n} \psi_{2n-1} \ ,
\end{eqnarray}
which is large and of order $1/\sqrt{\hbar}$. Again, this coupling
is of order $\sqrt{N_c}\l^{3/2}$ after the book-keeping. This
contribution is similar to (\ref{U1source}) apart from the
$SU(2)$ labels.

The isospin current is,
\begin{eqnarray}
J^\m_{I,a}(x)  =  -\sum_{n,m}   m_{v^n}^2 a_{{v}^n} \psi_{2n-1}
\int d^4{y}\ \  \k K \mathbb{F}^a_{Z \n}(y,Z)
\Delta_{mn}^{\n\m}(y-x) \Big|_{Z=B}
 \ ,
  \label{IVcurrent}
\end{eqnarray}
From (\ref{Bcurrent2}) and (\ref{IVcurrent}) the electromagnetic
current is given by
\begin{eqnarray}
J^\m_{\mathrm{EM}}(x) &=& J^\m_{I,3}(x) + \half J^\m_{B}(x) \nn \\
&=& -\sum_{n,m}   m_{v^n}^2 a_{{v}^n} \psi_{2m-1} \int d^4{y}\ \
\calq_\n(y,Z) \Delta_{mn}^{\n\m}(y-x) \Big|_{Z=B} \ ,
\label{EMcurrent}
\end{eqnarray}
with
\begin{eqnarray}
\calq_{\m}(x,Z) \equiv  \k  K {\mathbb{F}^3}_{Z\m}({x},Z)   +
\frac{1}{N_c} \k K \widehat{\mathbb{F}}_{Z\m}({x},Z) \ .
\end{eqnarray}
The electromagnetic charge density is
\begin{eqnarray}
J^0_{\mathrm{EM}}(x) &=&  -\sum_{n}   \int d^4{y}\ \calq_0(y,Z)
\int \frac{d^4{p}}{(2\pi)^4}
  e^{-i {p}\cdot ({y}-{x})} \frac{ {m}_{v^n}^{\ 2} }{{p}^2 +
  m_{v^n}^2}  a_{{v}^n} \psi_{2n-1}  \Big|_{Z=B} \nn \\
  &&+ \sum_{n}   \int d^4{y}\ \calq_{\m}(y,Z)
\int \frac{d^4{p}}{(2\pi)^4}
  e^{-i {p}\cdot ({y}-{x})} \frac{ p^\m p^0 }{{p}^2 +
  m_{v^n}^2}  a_{{v}^n} \psi_{2n-1}  \Big|_{Z=B} \ ,
  \label{J0general}
\end{eqnarray}
and the static electromagnetic form factor follows readily in the
form
\begin{eqnarray}
J^0_{\mathrm{EM}}(\vec{q}) &=&  \int d \vec{x}
e^{i\vec{q}\cdot\vec{x}} J^0_{\mathrm{EM}}({x}) \nn \\
&=&  - \sum_{n}  \  \int dZ  \dell_Z \left[ \left( \int  d \vec{x}
e^{i\vec{q}\cdot \vec{x}}  \calq_0({x},Z)\right) \psi_{2n-1}
\right] \frac{a_{v^n} m_{v^n}^2}{\vec{q}^{\ 2} + m_{v^n}^2} \ ,
 \label{FF1}
\end{eqnarray}
after setting $p^0=0$ in (\ref{J0general}) so that $\calq_i$ is
irrelevant. Recall that the instanton configuration we are using
is adiabatically rotating in flavor space. In the charge
$\cal{Q}_0$, these rotations generate a velocity dependence to
leading order in $\hbar$ which is proportional to the angular
momentum upon semiclassical quantization. With this in mind,
there is effectively no time-dependence left in the density
${\cal Q}_\mu (x, Z)$ to leading order in $\hbar$.

We now note that the electromagnetic form factor (\ref{FF1}) can
be rewritten as
\begin{eqnarray}
J^0_{\mathrm{EM}}(\vec{q}) = \sum_n \left(g_{V,min}^n
+g_{V,mag}^n\right)\,\frac{  a_{v^n} m^2_{{v}^n}}{{\vec{q}^{\
2}}+ m_{v^n}^2 } \ , \label{FF2}
\end{eqnarray}
with
\begin{eqnarray}
&& g_{V,min}^n = \int dZ  \dell_Z \left[ \left( \int  d \vec{x}
e^{i\vec{q}\cdot \vec{x}}  \calq_0({x},Z)\right) \right]
\psi_{2n-1} \ , \nn \\
&& g_{V,mag}^n =
 \int dZ  \left( \int  d \vec{x}
e^{i\vec{q}\cdot \vec{x}}  \calq_0({x},Z)\right)  \dell_Z
\psi_{2n-1} \ , \nn
\end{eqnarray}
which are the analogue of the minimal and magnetic coupling used
in the effective baryon description of~\cite{RHO}. The solitonic
character of the solution implies that the two contributions are
{\it tied} and sum up to a purely surface term in the Z-direction
a point that is not enforced in the effective approach~\cite{RHO}.
Also our results are organized in $\hbar$ starting from the original
D4 instanton.

\section{Electromagnetic Charge and Charge Radius}

The nucleon electromagnetic form factor is  written as a boundary
term
\begin{eqnarray}
J^0_{\mathrm{EM}}(\vec{q}) &=& \int d \vec{x}
e^{i\vec{q}\cdot\vec{x}} J^0_{\mathrm{EM}}(\vec{x}) \nn \\
&=& \int d \vec{x} e^{i\vec{q}\cdot\vec{x}} \sum_n \frac{a_{{v}^n}
m_{v^n}^{\ 2}}{\vec{q}^{\ 2} + m_{v^n}^2} \
 \psi_{2n-1}(Z_C) 2 \calq_0(\vec{x},Z_C)  \ ,
\end{eqnarray}
where the boundary term at $Z=\infty$ vanishes since $\psi_{2n-1}
\sim 1/Z $ for large $Z$.  In the limit ${q} \ra 0$ we pick the
electromagnetic charge
\begin{eqnarray}
\int d \vec{x}\,e^{i\vec{q}\cdot\vec{x}}\,2 \calq_0(\vec{x},Z_C) \ ,
\end{eqnarray}
due to the sum rule (\ref{sumrule1}). Since $Z_c$ will be set to
zero ultimatly at large $\l$, the limits $\lim_{q\ra 0} \lim_{Z\ra 0}$
will be understood sequentially. To proceed, we need to work out the
surface densities
$\calq_0$ i.e. the $U(1)$ and $SU(2)$ parts $
K{\widehat{\mathbb{F}}}_{Z0}(\vec{x},Z_c)$ and $ K
{{\mathbb{F}}}_{Z0}(\vec{x},Z_c)$ respectively. By the equations
of motion ($\ref{EOM:Full:4}$) and ($\ref{EOM:Full:1}$), they read
\begin{eqnarray}
&& \frac{4}{N_c} \k K {\widehat{\mathbb{F}}}_{Z0}(Z_c) =
\int_{-Z_C}^{Z_C} dZ \frac{1}{32 \pi^2}\e_{MNPQ} \left(\Tr
(\mathbb{F}_{MN}\mathbb{F}_{PQ}) +  \half
\widehat{\mathbb{F}}_{MN} \widehat{\mathbb{F}}_{PQ} \right) \nn \\
&& \qquad  \qquad \qquad \quad + \frac{2}{N_c}\int_{-Z_C}^{Z_C}
dZ \k K^{-1/3} \dell^i \widehat{\mathbb{F}}_{0i}  \ ,
\label{F0Zsol0}  \\
&& 2 \k K {{\mathbb{F}^a}}_{Z0}(Z_c) = \int_{-Z_C}^{Z_C} dZ \ 2i
\k \ \Tr \left\{ K^{-1/3} \left( [ \mathbb{F}_{0i},\mathbb{A}_i]
+ K
[ \mathbb{F}_{0Z},\mathbb{A}_Z] \right) t^a \right\} \nn \\
&& \qquad  \qquad \qquad \quad + \int_{-Z_C}^{Z_C} dZ \k K^{-1/3}
\dell^i \mathbb{\mathbb{F}}^a_{0i} + \int_{-Z_C}^{Z_C} dZ
\frac{N_c}{64\pi^2}\e_{MNPQ}( \wh{\mathbb{F}}_{MN}
\mathbb{F}_{PQ}^a) .
\end{eqnarray}

The $U(1)$ number density readily integrates to 1 since
\begin{eqnarray}
B=\int d\vec{x} J^0_{B }(\vec{x}) = \int d\vec{x} \frac{4}{N_c} \k K
{\widehat{\mathbb{F}}}_{Z0}(Z_c) = \int d\vec{x}\int_{-Z_C}^{Z_C}
dZ \frac{1}{32\pi^2}\e_{MNPQ} \Tr
(\mathbb{F}_{MN}\mathbb{F}_{PQ}) = 1 \ ,
\end{eqnarray}
as the spatial flux vanishes on $R_X^3$ and the $U(1)$ winding
number arezero for a sufficiently localized SU(2) instanton in
$R_X^3\times R_Z$. We note that the integrand is manifestly
gauge invariant. To contrast, the isovector charge is
\begin{eqnarray}
I^A= \int d\vec{x} 2 \k K {{\mathbb{F}^A}}_{Z0}(Z_c) =
 \int d\vec{x}\int_{-Z_C}^{Z_C} \,dZ \ 2i \k \ \Tr \left\{ K^{-1/3} \left( [
\mathbb{F}_{0i},\mathbb{A}_i] + K [ \mathbb{F}_{0Z},\mathbb{A}_Z]
\right) t^A \right\}  \ , \label{ISO}
\end{eqnarray}
again after dropping the surface term in $R_X^3$ and the
Chern-Simons contribution for a sufficiently localized instanton
in $|Z_C|$. Although the integrand in (\ref{ISO}) is not
manifestly gauge invariant, it is only sensitive to a rigid gauge
transformation at $Z_C$ which is reabsorbed by gauge rotating
the cloud as is explicit from the mode decomposition.

As noted earlier, the D4 instanton has maximal spherical symmetry
so that its isospin $I^A$ is just minus its angular momentum $J^A$,
\begin{eqnarray}
J^A &=& \int d\vec{x}dZ J^{0A} = \int d\vec{x}dZ \e_{Ajk}x^jT^{0k}
\nn \\
&=& \int d\vec{x} dZ \e_{Ajk}x^j \left[-\k
K^{-1/3}\mathbb{F}^{l0,a}\mathbb{F}^{lk,a} - \k K
\mathbb{F}^{Z0,a}\mathbb{F}^{Zk,a} \right] \ . \label{ANGULAR}
\end{eqnarray}
Both $I^A$ and $J^A$ are driven by the adiabatic rotation
$\mathbb{R}$. For the D4 instanton part, it is
\begin{eqnarray}
\mathbb{A}_R^a = \mathbb{R}^{ab}(t) \mathbb{A}^b \ , \qquad
\dot{\mathbb{A}}^a_R = \left(\dot{\mathbb{R}}(t)
\mathbb{R}^{-1}(t)\right)^{ab} \mathbb{A}_R^b \ , \label{ROT}
\end{eqnarray}
with $\w^A G^A \equiv -\dot{\mathbb{R}}\mathbb{R}^{-1}$~
\footnote{We recall that both the BPST instanton and the
fluctuations are rotating in the body fixed frame. As noted
earlier, the R-labeling of the fields is subsumed.}.
The $\omega$'s are quantum and of order $\hbar$.
Recalling the result for $\mathbb{A}_0^R$ for the constrained
field and the zero mode ($\mathbb{Z}^R$) from Appendix A,
we obtain to leading order
\begin{eqnarray}
J^A=-I^A &=& \int d\vec{x} dZ \e_{Ajk}x^j \left[-\k
K^{-1/3}\mathbb{F}^{l0,a}\mathbb{F}^{lk,a} - \k K
\mathbb{F}^{Z0,a}\mathbb{F}^{Zk,a} \right]\nn \\
&=&
\int d\vec{x} \int_{-Z_c}^{Z_c}dZ \e_{Ajk}x^j \left[-\k
(\mathbb{D}^{M}\mathbb{Z}^R)^a\mathbb{F}^{Mk,a}  \right]\nn \\
&\rightarrow& -\frac 12 M_0\rho^2 \,\w^A \equiv - \mathbb{I}\,\omega^A \ .
\end{eqnarray}
The last relation follows from the BPST instanton (\ref{ADHM3}).
As expected, the core instanton has a moment of inertia
$\mathbb{I}=M_0\rho^2/2$ where $M_0=8\pi^2\k$ is the D4 instanton
mass in units of $M_{KK}$ and to leading order in $1/\l$.
$\mathbb{I}$ is of order $N_c$.  Maximum spherical symmetry
results in a symmetric inertia tensor.

Finally, the nucleon charge is then
\begin{eqnarray}
&& \int d \vec{x} J^0_{\mathrm{EM}}(x) =  \int d \vec{x} 2
\calq_0(x,Z_c) = I_3 + \frac{B}2 \ .
\end{eqnarray}
While our analysis is to order $\hbar^0$ this normalization
should hold to all orders in $\hbar$. The vector meson cloud
encodes the exact charges in holography thanks to the exact sum
rule (\ref{sumrule1}).

The electromagnetic charge radius
$\langle r^2 \rangle_{EM}$ can be read from the
$q^2$ terms of the form factor
\begin{eqnarray}
\langle r^2 \rangle_{EM} =  \int d\vec{x}\ r^2 \,2
\calq_0(\vec{x},Z_c) + 6 \sum_{n=1}^{\infty} \frac{\a_{v^n}
\psi_{2n-1}(Z_c)}{m_n^2} \int d\vec{x}\ 2\calq_0(\vec{x},Z_c) \ ,
\end{eqnarray}
with $r \equiv \sqrt{(\vec{x})^2}$.  The first contribution is
from the core, while the second contribution is from the cloud.
For a sufficiently localized instanton in bulk the first
contribution is of order $1/\l$,
\begin{eqnarray}
\langle r^2 \rangle_{I=0}'&=&\int d\vec{x}\ r^2 \,2 \calq_0(\vec{x},Z_c)\nn \\
&=& \frac{3}{2}\rho^2\frac{Z_c}{\sqrt{Z_c^2+\rho^2}} \rightarrow \frac 32 \rho^2\, .
\end{eqnarray}
The meson cloud contribution is of order $\l^0$. It can be exactly asessed
by noting that
\begin{eqnarray}
\sum_{n=1}^{\infty} \frac{\a_{v^n} \psi_{2n-1}(Z_c)}{m_n^2} &=&
\int dZ  \sum_{n=1}^{\infty} \frac{\psi_{2n-1}(Z_c)
\psi_{2n-1}(Z) K^{-1/3}(Z)}{m_n^2} \nn \\
&=&   \int dZ \bra{Z_c}\Box_{\mathbf{C}}^{-1}\ket{Z}  \ ,
\end{eqnarray}
where $\Box_\mathbf{C}^{-1} \equiv -\dell_Z^{-1}K^{-1}
\dell_Z^{-1} K^{-1/3} $ is the inverse of
(\ref{EigenEqn}). This is just the vector meson propagator in curved
space~\footnote{The vector-meson coupling to the
instanton in the propagator is subleading in $1/\l$ and thus
dropped.}. It follows that
\begin{eqnarray}
\sum_{n=1}^{\infty} \frac{\a_{v^n} \psi_{2n-1}(Z_c)}{m_n^2} &=& -
\int dZ \int dZ'
\bra{Z_c}\dell_Z^{-1}\ket{Z'}K^{-1}(Z')\bra{Z'}\dell_Z^{-1}\ket{Z}K^{-1/3}(Z)
\ , \label{Integ}
\end{eqnarray}
where $\bra{Z'}\dell_Z^{-1}\ket{Z} = \half \mathrm{sgn}(Z' - Z)$
and $K=1+Z^2$. It is zero for $Z_c=\infty$ and $2.377$ for
$Z_c=\tilde{Z}_c/\sqrt{\l}$ in the double limit
$\l\rightarrow\infty$ followed by $\tilde{Z}\rightarrow\infty$.
See Appendix C for details. Thus the charge radius for the
nucleon is
\begin{eqnarray}
\sqrt{\langle r^2 \rangle_{EM}} \approx 14.26 \left(\half +
I_3\right) \Mkk^{-2} \approx 0.784 \left(\half + I_3\right)
\mathrm{fm}  ,
\end{eqnarray}
where we used $\Mkk = 950 \mathrm{MeV}$ \cite{Sakai1}. The
experimental values are \cite{Yao}
\begin{eqnarray}
\sqrt{\langle r^2 \rangle_{EM}^{\mathrm{proton}}} = 0.875\
\mathrm{fm} \ , \qquad \langle r^2
\rangle_{EM}^{\mathrm{neutron}} = -0.1161\ \mathrm{fm}^2 \ .
\end{eqnarray}

\section{Baryon Magnetic moment}

The magnetic moments follow from the moments of the electromagnetic
current,
\begin{eqnarray}
\mu^i &=& \half  \e_{ijk} \int d\vec{x} x^{j} J^k_{\mathrm{EM}}(x) \nn \\
&=&  \e_{ijk}\int d \vec{y} \sum_{n} m_{v^n}^2 a_{{v}^n}
\psi_{2n-1}(Z_c)   \calq_m(\vec{y},Z_c) \int d\vec{x}  x^{j}
\Delta_{n}^{mk}(\vec{y}-\vec{x}) \nn \\
&=& \e_{ijk}\int d \vec{y} \sum_{n} m_{v^n}^2 a_{{v}^n}
\psi_{2n-1}(Z_c)   \calq_m(\vec{y},Z_c) y^{j}
\frac{-g^{mk}}{m_{v^n}^2} \nn \\
&=& - \e_{ijk}\int d \vec{y} y^j \calq^k(\vec{y},Z_c) \ ,
\end{eqnarray}
with
\begin{eqnarray}
&&\calq_k(\vec{x},Z_c) \equiv  \k  K(Z_c)
{\mathbb{F}^3}_{Zk}(\vec{x},Z_c) + \frac{1}{N_c} \k K(Z_c)
\widehat{\mathbb{F}}_{Zk}(\vec{x},Z_c) \ .
\label{SOURCE}
\end{eqnarray}
While the electromagnetic current is meson mediated at the core
boundary, its contribution to the magnetic moment to order
$\hbar^0$ is core-like owing to the exact sum-rule (\ref{SUM}) in
warped space. By resumming over the tower of infinite vector
mesons, the magnetic moment shrunk to the core at strong coupling
with $g^2$ large and $N_c$ large. This remarkable feature is
absent in the Skyrme model and its variants since they all
truncate the number of mesons.

First consider the iso-scalar contribution to the magnetic moment
in (\ref{SOURCE}). As we are assessing $\calq_k(x,Z_c)$ at the core
boundary, the small $Z$ instanton configuration is sufficient.
From (\ref{EOM:Full:5}) it follows that
\begin{eqnarray}
\frac{1}{N_c}  \k {{\widehat{\mathbb{F}}}}_{Z k}^R(\vec{x},Z_c)
&=& \int_{-Z_c}^{Z_c} dZ \left[ \frac{1}{32\pi^2}\e_{kNPQ}\Big(
\Tr({\mathbb{F}}_{N0}^R {\mathbb{F}}_{PQ}^R) \right]
\nn \\
 &=& \int_{-Z_c}^{Z_c} dZ \left[
\frac{1}{32\pi^2}\e_{kNPQ}\Big( \Tr(\mathbb{D}_n \mathbb{Z}^R
{\mathbb{F}}_{PQ}) \right] , \label{SCALAR}
\end{eqnarray}
where we have dropped the U(1) CS contribution as it is subleading
as well as surface contributions on $R_X^3$ since we will
integrate over $R_X^3$ at the end.  The upper R-labels refer to
the rigid SO(3) rotation $\mathbb{R}$ and $\mathbb{Z}^R$ is the
zero mode from the Gauss constraint (Appendix A).
In the second line the R-label drops because of tracing. Thus
\begin{eqnarray}
\m^A_{I=0} &=&  - \e_{Ajk}\int d \vec{y} y^j \frac{2}{N_c} \k
K(Z_c)
\widehat{\mathbb{F}}_{Zk}^R(\vec{x},Z_c) \nn \\
&=& -\frac{\rho^2 \w^A}{4} \frac{Z_c}{\sqrt{Z_c^2 + \rho^2}} \rightarrow -\frac{\rho^2 \w^A}{4} \nn \\
&=&\frac{\langle r^2 \rangle_{I=0}'}6\,\frac{J^A}{\mathbb{I}}
 \ ,
\end{eqnarray}
where we used the BPST solution (\ref{ADHM3}) since
$Z_c\sim\rho\sim 0$. The contribution is of order $\hbar$ but
subleading in $1/\lambda$. The last relation uses that $I^A=-J^A$.
This relation for the isoscalar magnetic moment is similar to the
one derived in the Skyrme model with the notable difference that
only the isoscalar core radius and core moment of inertia are
involved.

Now, consider the iso-vector contribution to the magnetic moment.
Using (\ref{EOM:Full:2}), we have
\begin{eqnarray}
\k {{\mathbb{F}}}_{Z k}^{R,3}(\vec{x},Z_c)   &=& \int_{-Z_c}^{Z_c}
dZ i \k \Tr
\left\{  [{\mathbb{A}}^{R,M}, {{\mathbb{F}}}_{M k}^{R}] t^3 \right\} \nn \\
 &\rightarrow&
-\int_{-Z_c}^{Z_c} dZ \frac{\k}{4} \mathbb{A}_M^a
\mathbb{F}_{Mk}^b \e_{abc}\mathbb{R}_{3c} \ .
\end{eqnarray}
Much like the iso-scalar, we have only retained the leading
contribution to the magnetic moment. As noted earlier, the residual
gauge variance of the integrand through $Z_C$ is removed by gauge rotation of the
cloud. In terms of the regular BPST solution (\ref{ADHM2}) and (\ref{ADHM3}) we get
\begin{eqnarray}
 \mathbb{A}_M^a \mathbb{F}_{Mk}^b \e_{abc} =
 \frac{-8\rho^2}{(\xi^2+\rho^2)^3}(x_a\e_{akc} - x_b\e_{kbc}) \ ,
\end{eqnarray}
so that
\begin{eqnarray}
\m^i_{I=1} &=&   - \e_{ijk}\int d \vec{y} y^j  \k  K(Z_c)
{\mathbb{F}^{R,3}}_{Zk}({x},Z_c) \nn \\
&=& -\frac{32\pi}{3} \k \rho^2 \mathbb{R}_{3i} \int_0^\infty dr
\int_{-Z_c}^{Z_c} dZ \frac{r^4}{(\xi^2+\rho^2)^3} \nn \\
&=& -{4\pi^2 }  \k \rho^2 \mathbb{R}_{3i} \log_c \nn \\
&=&-\frac{\mathbb{I}}2\,{\mathbb{R}}^{3i}\log_c \, ,
\end{eqnarray}
with a logarithmic cutoff sensitivity to the core size,
\begin{eqnarray}
\log_c \equiv \log\left( \frac{ Z_c+\sqrt{Z_c^2 +
\rho^2}}{-Z_c+\sqrt{Z_c^2 + \rho^2}} \right)\,\,.
\end{eqnarray}
The isovector magnetic contribution is of order $\hbar^0$ and similar
in structure to the Skyrmion, with the exception that $\mathbb{I}$ is
solely driven by the core. The cutoff sensitivity $\log_c$ is absent if
we were to use the BPST instanton in the singular gauge. The Cheshire
Cat smile survives in the isovector magnetic contribution in the
regular gauge (albeit weakly through a logarithm).

Combining the isoscalar and isovector contributions to the magnetic
moment yields (singular gauge)
\begin{eqnarray}
\mu^i= \frac{\langle r^2 \rangle_{I=0}'}6\,\frac{J^i}{\mathbb{I}}
-\frac{\mathbb{I}}2\,{\mathbb{R}}^{3i} \ ,
\end{eqnarray}
which results in the Skyrme-like independent relation~\cite{SKYRME},
\begin{eqnarray}
\frac{\mu_p-\mu_n}{\mu_p+\mu_n} =\frac 34
\frac{M_\Delta+M_N}{M_\Delta-M_N} \ ,
\end{eqnarray}
expected from a soliton. Here $M_{N,\Delta}$ are the nucleon and
delta masses split by the inertia $\mathbb{I}$.

\section{Conclusions}

We have shown how the non-rigid quantization of the D4 instanton
in holographic QCD yields baryon electromagnetic form factors that
obey the strictures of VMD in agreement with the effective
approach discussed in~\cite{RHO}. The  holographic baryon at the
boundary is composed by a core instanton in the holographic
direction at $Z=0$ of size $1/\sqrt{\l}$ that is trailed by a
cloud of bulk vector mesons and pions of size $\lambda^0$ all the
way to $Z=\infty$. The core and the cloud interface at $Z_C$
which plays the role of the Cheshire Cat smile (gauge movable).
At strong coupling, the baryon size is of order $\lambda^0$
thanks to vector meson dominance. The meson-baryon couplings are
large and of order $1/\sqrt{\hbar}$ (or lower) and surface-like
only owing to the solitonic nature of the instanton.

The electromagnetic form factors, radii and magnetic moments of
the ensuing baryons compare favorably with the results obtained
in the Skyrme model, as well as data for a conservative value of
$M_{KK}=950$ MeV. For instance the electromagnetic charge radii are
$0.784\ \mathrm{fm}$ (proton) and zero (neutron). They are derived in the
triple limit of zero pion mass (chiral limit) and strong coupling $\l$
(large $g^2$ and large $N_c$). The magnetic moments are completly
driven by the core D4 instanton through a remarkable sum rule of
the vector meson cloud. They obey a model independent relation
of the type encountered in the Skyrme model, a hall-mark of large
$N_c$ and strongly coupled models.

The non-rigid quantization scheme presented here offers a
systematic framework for discussing quantum baryons in the
context of the semi-classical approximation. It is causal
with retardation effects occuring in higher order in $\hbar$.
These semiclassical corrections are only part of a slew of
other quantum corrections in holography which are in
contrast hard to quantify. Also, the small instanton size
calls for the use of the full DBI action to characterize
the instanton field more faithfully in the holographic core.
We note that beyond the $\hbar^0$ contribution discussed
here, the issue of Dirac constraints needs to be addressed.
This is best addressed in the canonical Hamiltonian formalism
whereby Gauss laws are explicitly removed by their constaint
equations. The drawback are lack of manifest covariance and
operator orderings.

The extension of our analysis to the axial-vector channels is
straightforward with minimal changes in our formulae as can be
readily seen by inspection. Indeed, the axial vector source
differs from the vector one we used by the extra mode function
$\psi_0(Z)$ which is odd in Z. Also the pion field $\Pi$ now
contributes. In the axial-vector channel the pion-baryon coupling
is expected to be formally of order $\sqrt{\hbar}$ but {\it
time-like} thus effectively of order $1/\sqrt{N_c}$ and not
$\sqrt{N_c}$. The Goldberger-Treiman relation in this case
follows from the non-rigid quantization of the instanton much
like its counterpart for the Skyrmion~\cite{VERSCHELDE}. Some of
these issues and others will be discussed next.

\section{Acknowledgments}
We would like to thank S.~J. Sin for valuable discussion. K.Y.K.
is grateful to H.U. Yee for useful discussions. This work
was supported in part by US-DOE grants DE-FG02-88ER40388 and
DE-FG03-97ER4014.

\vskip 2cm

{\bf Note Added:} Upon completion of this work we noted
arXiv:0806.3122 \cite{Hashimoto} where similar issues are
addressed with different methods.

\appendix

\section{Gauss Law}

For the flavor rotated instanton, the rotated form of Gauss law
(\ref{EOM:Full:1} ) reads
\begin{eqnarray}
\mathbb{D}^R_M{\mathbb{F}}^R_{M0}={\cal O}\left(1/\lambda\right)
\ ,
\end{eqnarray}
to leading order as the curvature effects are subleading in
$1/\lambda$. All upper R-labels refer to the rigid SO(3) rotation
$\mathbb{R}$ with $\mathbb{D}^R_M=\partial_M+\mathbb{A}^{R,A}G^A$.
While this is subsumed throughout in the main text, here it is
recalled explicitly for the clarity of the argument. The formal
solution reads
\begin{eqnarray}
\mathbb{A}_0^R=\frac 1{(\mathbb{D}^{R'}
)^2}\,\mathbb{D}^R_N\mathbb{\dot{A}}_N^R+\mathbb{Z}^R \ ,
\label{SOL}
\end{eqnarray}
where the primed inverted operator excludes the zero mode. Following
\cite{Sakai3}, the rotated zero mode solution reads
\begin{eqnarray}
\mathbb{Z}^{R,A}=C\,\mathbb{R}\mathbb{\overline{R}}(g)\omega^A
\,f(\xi) \ ,
\end{eqnarray}
up to an arbitrary constant $C$. Here $f(\xi)=\xi^2/(\xi^2+\rho^2)$
and
\begin{eqnarray}
\overline{\mathbb{R}}^{AB}(g)= \Tr \left(t^Ag^{-1}t^Bg\right) \ ,
\qquad\qquad g= \frac{z-i\vec{x}\cdot\vec{\s}}{\xi} \ .
\end{eqnarray}
We note that for an unrotated BPST instanton with $\omega^A=0$,
the formal solution (\ref{SOL}) yields $\mathbb{A}_0=0$ as it
should. The normalization is $C=1$ which is fixed by the
asymptotic of the $\mathbb{A}^R_0$ field:
$\mathbb{A}_0^R(x,Z=\infty) = U^{R\dagger}\dell_0 U^R$.
$U=e^{i\t_a\hat{r}^a(\theta,\phi,\psi)}$ is the identity map and
$(\theta,\phi,\psi)$ are the canonical angles on $S^3$.

In terms of (\ref{SOL}) the rotated electric field is
\begin{eqnarray}
\mathbb{F}_{M0}^R=\mathbb{D}_M^R\mathbb{A}_0^R-\dot{\mathbb{{A}}}_M^R=\mathbb{D}_M^R\mathbb{Z}^R\,\,.
\end{eqnarray}
The kinetic energy for the rotated instanton is
\begin{eqnarray}
H= {\kappa}\int d\vec{x}\,dZ\,\Tr (\mathbb{F}_{M0}^R)^2 =  \kappa
\int d\vec{x}\,dZ\, \Tr \left(\mathbb{D}_M \mathbb{Z}^R\right)^2
=\frac 12 M_0\rho^2 \w^2 \ .
\end{eqnarray}
The upper R-label drops out by tracing. Here $M_0=8\pi^2\kappa$
is the instanton mass in units of $M_{KK}$ to leading order in $1/\l$.

\section{Action}

With the gauge field (\ref{GF1})-(\ref{GF4}) the $ SU(2)$ YM
action reads
\begin{eqnarray}
&& S_{YM}^{SU(2)} = -\k  \int d^4x dZ  \Tr  \left[ \half K^{-1/3}
\mathbb{F}_{\m\n}^2    + K \mathbb{F}_{Z\m}^2  \right. \nn \\
&& \qquad \qquad\qquad\qquad\qquad + \dell^\m
\left(2K^{-1/3}\mathbb{F}_{\m\n} C^\n\right) + \dell^Z \left(2K
\mathbb{F}_{Z\n}C^\n\right) + \dell^\m \left(2K
\mathbb{F}_{\m Z}C^Z\right)  \nn \\
&& \qquad \qquad\qquad\qquad\qquad  - \left\{ \mathbb{D}^\m
\left(2 K^{-1/3} \mathbb{F_{\m\n}} \right) +  \mathbb{D}^Z \left(
2K \mathbb{F_{Z\n}} \right) \right\} C^\n -   \mathbb{D}^\m \left(
2 K
\mathbb{F_{\m Z}} \right) C^Z  \nn \\
&& \qquad \qquad\qquad\qquad\qquad   +  \half K^{-1/3}
\left\{2\mathbb{F}_{\m\n} [C^\m, C^\n] + \Big( \mathbb{D}_\m C_\n
- \mathbb{D}_\n C_\m -i [C_\m, C_\n]\Big)^2
\right\} \nn \\
&& \qquad \qquad\qquad\qquad\qquad  \left. + K
\left\{2\mathbb{F}_{Z\m} [C^Z, C^\m] + \Big( \mathbb{D}_Z C_\m -
\mathbb{D}_\m C_Z -i [C_Z, C_\m]\Big)^2 \right\} \right]
\label{SU2} \ ,
\end{eqnarray}
where $ \mathbb{D}_\a $ is the covariant derivative with the
soliton configuration: $\mathbb{D}_\a * = \dell_\a
-i[\mathbb{A}_\a , *]$.  Similary U(1) YM action is
\begin{eqnarray}
&&S_{YM}^{U(1)} =  - \frac{\k}{2} \int d^4x dZ    \left[ \half
K^{-1/3}
\widehat{\mathbb{F}}_{\m\n}^2    + K \widehat{\mathbb{F}}_{Z\m}^2  \right. \nn \\
&& \qquad \qquad\qquad\qquad\qquad + \dell^\m
\left(2K^{-1/3}\widehat{\mathbb{F}}_{\m\n} \widehat{C}^\n\right)
+ \dell^Z \left(2K
\widehat{\mathbb{F}}_{Z\n}\widehat{C}^\n\right) + \dell^\m
\left(2K
\widehat{\mathbb{F}}_{\m Z}\widehat{C}^Z\right)  \nn \\
&& \qquad \qquad\qquad\qquad\qquad  -  \left\{ \dell^\m  \left( 2
K^{-1/3} \widehat{\mathbb{F}}_{\m\n} \right) +  \dell^Z \left( 2 K
\widehat{\mathbb{F}}_{Z\n} \right) \right\} \widehat{C}^\n -
\dell^\m \left( 2 K
\widehat{\mathbb{F}}_{\m Z} \right) \widehat{C}^Z  \nn \\
&& \qquad \qquad\qquad\qquad\qquad \left.   +  \half K^{-1/3}
\Big( \dell_\m \widehat{C}_\n - \dell_\n \widehat{C}_\m \Big)^2 +
K  \Big( \dell_Z \widehat{C}_\m - \dell_\m \widehat{C}_Z \Big)^2
\right] \label{U1} \ ,
\end{eqnarray}
and the CS term is
\begin{eqnarray}
S_{CS} &=&  \frac{N_c}{24\pi^2}\e_{MNPQ} \int d^4x dZ \nn
\\
&& \left[  \frac{3}{8} \Big(\wh{\mathbb{A}}_0 +
\widehat{C}_0\Big)\ \Tr \Big\{ \mathbb{F}_{MN}\mathbb{F}_{PQ}
+ 4\mathbb{F}_{MN} \mathbb{D}_P C_Q + 4\mathbb{F}_{MN} C_P C_Q )  \right. \nn \\
&& \qquad \qquad \qquad \quad + 4(\mathbb{D}_M C_N + C_M C_N)(\mathbb{D}_P C_Q + C_P C_Q ) \Big\} \nn \\
&& -\frac{3}{2}\Big( \wh{\mathbb{A}}_M  + \wh{C}_M \Big)\ \Tr
\Big\{\dell_0 (\mathbb{A}_N + C_M )
( \mathbb{F}_{PQ} + 2\mathbb{D}_P C_Q + 2 C_P C_Q   )\Big\} \nn \\
&& + \frac{3}{4} \Big( \wh{\mathbb{F}}_{MN} + 2\dell_M \wh{C}_N
\Big) \ \Tr \Big\{(\mathbb{A}_0 + C_0)
( \mathbb{F}_{PQ} + 2\mathbb{D}_P C_Q + 2 C_P C_Q   ) \Big\} \nn \\
&&+ \frac{1}{16} \Big( \wh{\mathbb{A}}_0 + \wh{C}_0 \Big) \Big\{
\wh{\mathbb{F}}_{MN}\wh{\mathbb{F}}_{PQ}
 +4\widehat{\mathbb{F}}_{MN}\dell_P \wh{C}_Q  + 4(\dell_M \wh{C}_N)( \dell_P \wh{C}_Q )\Big\}  \nn \\
&&- \frac{1}{4} \Big(\wh{A}_M + \wh{C}_M \Big)\Big\{
\wh{\mathbb{F}}_{0N}\wh{\mathbb{F}}_{PQ}
+2\wh{F}_{0N}\dell_P\wh{C}_Q + \wh{F}_{PQ}(\dell_0\wh{C}_N -
\dell_N \wh{C}_0 )  \nn \\
&&\qquad \qquad \qquad \quad + 2 (\dell_0 \wh{C}_N - \dell_N
\wh{C}_0)\dell_P\wh{C}_Q \Big\}
\nn \\
&& \left. + \frac{3}{2}\dell_N \left[  \Big( \wh{\mathbb{A}}_M +
\wh{C}_M \Big)  \Tr \Big\{ ( \mathbb{A}_0+C_0) ( \mathbb{F}_{PQ}
+ 2\mathbb{D}_P C_Q + 2C_P C_Q  ) \Big\} \right] \right] \nn \\
&&+ \frac{N_c}{48\pi^2} \int  \ud \left[ \Big(\widehat{\mathbb{A}}
+ \widehat{C}\Big)\ \Tr \left\{2 \ud (\mathbb{A} + C)(\mathbb{A}
+ C) - \frac{3i}{2} (\mathbb{A}+C)^3 \right\} \right] \ .
\label{CSex}
\end{eqnarray}
%
%


\section{Integral}

In this appendix we work out the integral in (\ref{Integ}):
\begin{eqnarray}
G(Z_c) \equiv - \int dZ \int dZ'
\bra{Z_c}\dell_Z^{-1}\ket{Z'}K^{-1}(Z')\bra{Z'}\dell_Z^{-1}\ket{Z}K^{-1/3}(Z)
\ . \label{G}
\end{eqnarray}
We start with the Green function
\begin{eqnarray}
\bra{Z'}\dell_Z^{-1}\ket{Z} = \half \mathrm{sgn}(Z'-Z) \ .
\end{eqnarray}
Since there are two sgn functions in ($\ref{G}$) we divide the
integral region into six pieces reflecting all possible sign
difference:
\begin{eqnarray}
G(Z_c) &=& - \frac{1}{4}\left[ \int_{-\infty}^{Z_c} dZ'
\int_{-\infty}^{Z'}dZ -
\int_{-\infty}^{Z_c} dZ' \int_{Z'}^{Z_c} dZ  \right. \nn \\
&&- \int^{\infty}_{Z_c} dZ' \int_{Z_c}^{Z'}dZ +
\int^{\infty}_{Z_c}
dZ' \int_{Z'}^{\infty} dZ \nn \\
&& \left. -\int_{Z_c}^{\infty}dZ'  \int_{-\infty}^{Z_c} dZ  -
\int^{Z_c}_{-\infty} dZ' \int^{\infty}_{Z_c} dZ
\right](K^{-1}(Z')K^{-1/3}(Z)) \ .
\end{eqnarray}
In two extreme(symmetric) case the expression becomes simple. For
$Z_c=\infty$
\begin{eqnarray}
-\frac{1}{4}\left[\int_{-\infty}^{\infty} dZ'
\int_{-\infty}^{Z'}dZ - \int_{-\infty}^{\infty} dZ'
\int_{Z'}^{\infty} dZ\right] (K^{-1}(Z')K^{-1/3}(Z)) = 0 \ ,
\end{eqnarray}
and for $Z_c=0$
\begin{eqnarray}
&& - \frac{1}{2} \left[ \int_{-\infty}^{0} dZ'
\int_{-\infty}^{Z'}dZ - \int_{-\infty}^{0} dZ' \int_{Z'}^{0} dZ
-\int_{0}^{\infty}dZ'
\int_{-\infty}^{0} dZ  \right](K^{-1}(Z')K^{-1/3}(Z)) \nn \\
&& =  \int_{0}^{\infty}dZ' K^{-1}(Z') \int_{0}^{Z'} dZ K^{-1/3}(Z)
\sim 2.377 \ .
\end{eqnarray}

\end{document}